# Generic nano-imprint process for fabrication of nanowire arrays

Aurelie Pierret[1], Moïra Hocevar[1,2], Silke L. Diedenhofen[3], Rienk E. Algra[1,4,5], E. Vlieg[5], Eugene C. Timmering[1], Marc A. Verschuuren[1], George W.G. Immink[1,6], Marcel A. Verheijen[1,6], Erik P.A.M. Bakkers[1]

[1]Philips Research Laboratories Eindhoven, High Tech Campus 11, 5656AE Eindhoven, The Netherlands

[2]Delft University of Technology, 2628 CJ Delft, The Netherlands

[3]FOM Institute for Atomic and Molecular Physics c/o Philips Research Laboratories High Tech Campus 4 5656 AE Eindhoven, The Netherlands

[4]Materials innovation institute (M2i), 2628CD Delft, The Netherlands

[5]IMM, Solid State Chemistry, Radboud University Nijmegen, Heyendaalseweg 135, 6525AJ Nijmegen, The Netherlands

[6]MiPlaza Technology Laboratories, Philips Research Europe, High Tech Campus 29 5656 AE Eindhoven, The Netherlands

*E-mail: erik.bakkers@philips.com

**Abstract**

A generic process has been developed to grow nearly defect free arrays of (heterostructured) InP and GaP nanowires. Soft nanoimprint lithography has been used to pattern gold particle arrays on full 2 inch substrates. After lift-off organic residues remain on the surface, which induce the growth of additional undesired nanowires. We show that cleaning of the samples before growth with piranha solution in combination with a thermal anneal at 550 °C for InP and 700 °C for GaP results in uniform nanowire arrays with 1% variation in nanowire length, and without undesired extra nanowires. Our chemical cleaning procedure is applicable to other lithographic techniques such as e-beam lithography, and therefore represents a generic process.

# 1. Introduction

Semiconductor nanowires (NWs) have been intensely studied because of their highly tuneable optical and electrical properties. They have been employed as functional elements in light emitting diodes (LEDs) [1-2], transistors [3], and antireflection coatings [4]. The Vapor-Liquid-Solid (VLS) bottom-up growth process [5] allows growing (radial) core/shell structures or (axial) quantum dot heterostructures [6]. A metal nanoparticle is used as catalyst for nanowire growth, and can be deposited via spin coating from a colloidal solution, spraying an aerosol solution, or by deposition of a thin film, which will break up by heating the sample. However, the density and position of the nanowires are not controlled using these approaches. We have previously observed [7] that the nanowire growth rate sensitively depends on the density and the dimensions of the gold particles. Such a variation of the growth rate can result in non uniform optical or electrical properties. It is likely that, for instance, the dispersion of the electron mobility in InAs/InP core-shell nanowires [8], is related to the large spread in shell thickness of these nanowires grown from randomly deposited colloidal gold particles. Besides the axial growth rate, most probably also the radial (shell) growth is affected by the catalyst density. Hence, in order to improve the nanowire uniformity, it is important to control the position and density of the catalyst particles on the substrate. Different approaches have already been reported, such as e-beam lithography [9-11], gold deposition through an anodic aluminium oxide template [12] and nanosphere lithography [13]. In general, these techniques do not allow for large-area structuring because of either high cost or lack of long range order. Alternatively, nanoimprint lithography [14] has been reported to define nanowire positions. This technique enables patterning of large surface areas at relatively low-cost. However, this process has not been fully optimised and extra undesired nanowires were obtained. It has been shown that nanowire growth can be nucleated by organic molecules [15] and the undesired nanowires probably origin from organic residues from the photoresist.

We have developed and optimized a wafer scale soft nano-imprint method, called Substrate Conformal Imprint Lithography (SCIL) to control the position of InP and GaP nanowires. In order to fabricate uniform, defect free nanowire arrays we have systematically studied the effect on nanowire growth of chemical and thermal treatment of the substrate which contained gold catalyst particles. These treatments were focused on the removal of organic residues from the photoresist layer. The process has been further optimized by adjusting the gold layer thickness. Finally, the quality of the grown nanowire material has been evaluated by optical characterization on InAsP quantum dots in InP nanowires.

### 2. Experimental details

Our substrate conformal imprint lithography method makes use of a flexible patterned stamp made from Poly-Di-Methyl-Siloxane (PDMS) [14] which is molded from a silicon master pattern which contained arrays of holes and was fabricated using e-beam lithography. The composite stamp is molded from the master following the procedure described in [16,17] which yields a two layer soft PDMS stamp with protruding pillars. The developed nano-imprint and lift-off process has the following steps. A ~100 nm thick layer of Poly-Methyl-Meth-Acrylat (PMMA) is applied by spin coating on full 2 inch annealed (111)B InP or freshly etched (111)B GaP wafers, after which the wafers with the PMMA layer are baked on a hotplate at 150 °C for 15 minutes. Next, a precise amount of silica based sol-gel imprint resist is applied by spin coating over the PMMA layer [18]. Within two minutes after the spin coat process the stamp is applied in the still liquid sol-gel resist. The features in the stamp are filled by capillary forces with the sol-gel resist. The sol-gel reacts to form solid silica glass in about 20 minutes, where residual solvents and reaction products (alcohols and water) are absorbed in the PDMS rubber. Finally the stamp is carefully removed from the patterned silica layer by peeling. In this way the full 2" wafer area is patterned with an array of holes with a diameter of ~ 100 nm. The ~10 nm thick remaining residual sol-gel layer between the bottom of the features and the PMMA layer is removed by a $CF_4$ based reactive ion etch (RIE). Then oxygen RIE is used to transfer the pattern of the sol-gel layer into the underlying PMMA layer. The oxygen etch stops as the III-V material is reached and a short over etch is applied in order to remove any residual PMMA. A short etch in a 1% HF solution removes the formed Ga or In oxide, after which a thin (1-6 nm) gold layer is deposited onto the patterned sol-gel / PMMA pattern by perpendicular evaporation. The lift-off process is performed in acetone which dissolves the PMMA layer and releases the silica and gold metal which are suspended by PMMA. This leaves precisely placed gold dots of controlled diameter and thickness on the semiconductor material. In this work different stamps were used with 90 and 100 nm hole diameters on a corresponding pitch of 0.5 μm (tetragonal) and 1 μm (hexagonal).

The lift-off process leaves (organic) residues on the substrate besides the patterned metal particles. Various chemicals were tested to clean the substrate surface. We investigated the effect of fumic nitric acid ($HNO_3$), Piranha ($H_2SO_4$ : $H_2O_2$ : $H_2O$, 5:1:1), King's Water (HCl : $HNO_3$ : $H_2O$, 3:2:3) and (1 %) fluoric acid (HF) and bromine-methanol ($Br_2$ : MeOH, $1:10^3$ or $1:25*10^3$) on the NWs growth quality. After exposure to these chemicals at room temperature the samples were left in streaming deionized cold water for about 5 min.

The nanowires were synthesized in an Aixtron 200 Metal-Organic Vapour-Phase Epitaxy (MOVPE) reactor from Tri-Methyl-Gallium (TMGa) or Tri-Methyl-Indium (TMIn) for the group III elements and Phosphine ($PH_3$) for the group V-elements. For growth of an InAsP heterostructure, Arsine ($AsH_3$) was added. A thermal anneal was applied before nanowire growth for 10 minutes at different temperatures under a $PH_3$ flow. During the thermal anneal the gold droplets form a eutectic, Au-In or Au-Ga, depending on the substrate. The native oxide layer covering the samples, and residual organic materials are removed at annealing temperature exceeding 550 °C.

Scanning Electron Microscopy (SEM) images were obtained with a Philips XL 40 FEG system. NWs lengths and diameters were measured in these images by investigating 30 nanowires per array. The samples were studied with a Tecnai 300 keV TEM in both brightfield as well as in high-angle annular dark field (HAADF) mode. The chemical composition of the nanowires was studied using scanning TEM in combination with energy-dispersive X-ray analysis (EDX).

Microphotoluminescence measurements on InAsP heterostructure nanowires transferred on $SiO_2$ substrates were performed at 4.2 K. The nanowire quantum dots were excited with a 532 nm continuous wave laser focused to a spot size of 0.6 $\mu$m using a microscope objective with a numerical aperture NA = 0.85.

## 3. Results and Discussion

Figure 1(a) shows a SEM image of the surface after the lift-off process without further treatment. We can clearly see the high fidelity of the pattern transfer, as the Au particles are arranged in an ordered pattern. However, extra material is present around the Au islands. EDX measurements, carried out with the SEM, show that these contaminants are carbon based residues from the PMMA layer. The nanowires are grown by MOVPE on (111)B oriented InP and GaP substrates, on which the Au particle arrays were patterned using SCIL. The SEM image in Figure 1(b) is taken after InP nanowire growth. Besides nanowires grown from the predefined catalyst particles in a regular pattern, a lot of undesired thin nanowires are obtained. These extra nanowires arise if no or improper pre-treatment was used after the lift-off. This shows that growth of these extra nanowires is initiated by the organic material observed in figure 1(a). An extra cleaning step is necessary to remove the organics prior to growth, but it should not change the III-V substrate surface chemistry [19], since nanowire growth rates sensitively depend on the surface diffusion of the precursor molecules [7, 10]. We have investigated the effect of the following wet chemical treatments on the quality of the nanowire growth: *Fuming nitric acid*, an acidic etchant, oxidizes InP and GaP. *Piranha* is a mixture that contains a very strong oxidizer which can remove organic residues. *King's Water* is a combination of a powerful oxidizer and strong acid, and *$Br_2$/methanol* is an electrochemical etchant for InP and GaP. After exposure to one of these solutions or a combination of them the samples were left in streaming deionized cold water for about 5 min. The results are summarized for InP and GaP in table 1 and table 2, respectively.

InP samples treated with King's Water, $HNO_3$, and ($HNO_3$+HF) show a large number of undesired nanowires around the intended nanowires. During the longer treatments with King's Water, the gold has been removed and only undesired nanowires were grown (see supplementary information Figure S1). Samples exposed to a piranha solution at 20°C show a very low density of undesired nanowire, as shown in Figure 1c and 1d. (Almost) no undesired nanowires were obtained and all the nanowires grow at the predefined position. The overview SEM image in Figure 1d shows that the InP nanowire dimensions are uniform across the sample and that just a few nanowires are missing. The obtained InP nanowires have an interspacing of 500 nm, as defined by the mold, and a length of 1761 ± 19 nm which is only a spread of 1.1%. Furthermore the nanowires are tapered and have a top diameter of 95 ± 4 nm and base diameter of 171 ± 4 nm. Important to note is that all InP samples presented in the table have been thermally annealed at 550°C before growth. The effect of the anneal temperature will be discussed below.

GaP samples treated with Br$_2$/MeOH, (HNO$_3$ + Piranha), (Piranha+ HNO$_3$) or Piranha for 5 min show not only extra nanowires (due to inherent added contamination with each processing step) but also misplaced and/or missing nanowires. We believe that these chemicals can oxidize and/or etch the Au/GaP interface, which can result in removal of Au particles from the surface. We found that, similar to the results for InP wires, piranha treatment gives the best results (see supplementary information Figure S1), and the optimum treatment time for GaP wire growth is 1 minute. At shorter times more undesired NWs arise, but at longer times NWs tend to kink and some of them are missing, which is probably due to an overetching of the surface. The best results were obtained in combination with the higher anneal temperature of 700°C. We will now show that an extra thermal anneal step is necessary to optimise the growth.

We have studied the effect of annealing temperature on the nanowire growth. The annealing was done after piranha treatment, but prior to the growth. It is well-known that a thermal anneal at high enough temperatures can remove In$_2$O$_3$ or Ga$_2$O$_3$ from the substrate surface [20]. To our best knowledge it is not known if the photoresist residues can be removed or react with the metal oxides by thermal treatment. In Figure 2 the effect of annealing on the growth of InP nanowires is shown. Clearly, too low annealing temperatures, or no annealing at all results in undesired extra wire growth and irregular growth of the patterned nanowires (see Figure 2a and 2b). An annealing temperature of at least 550 °C proves to be sufficient for defect free InP wire growth (Figure 2c). At higher temperatures nanowires with a thicker base are obtained (Figure 2d). Probably the surface chemistry changes at these temperatures and promote the lateral growth at the base of the nanowires. Similar results were obtained for the growth of GaP nanowires (see supplementary figure S2). Whereas at an annealing temperature of 550 °C still undesired GaP nanowires are obtained, at a temperature of 700 °C the defect density is very low (see also table 2). The GaP nanowires show slight tapering, but do not have a thick base as observed for the InP nanowires when annealed at this temperature (see supplementary information Figure S2). Growth without chemical cleaning but with an anneal step prior to growth resulted in many undesired nanowires. This shows that the combination of piranha treatment with a sufficiently high anneal temperature is essential for both InP and GaP wire growth. The different optimum annealing temperatures for InP and GaP shows that the removal process is substrate dependent and suggests that the metal oxides play an important role in removing the organic residues.

To define the pattern in our process, we have used PMMA in direct contact with the substrate surface. The organic contaminants, from which the undesired nanowires grow, are probably a product from the PMMA layer. Our cleaning process effectively removes these residues and this cleaning process should also work successful for other lithographic techniques, such as e-beam, in which the same photo resist is used. This shows that we have found a generic method to clean the substrate surface after lithography and to avoid the growth of undesired nanowires.

Now that we managed to control the NWs position on a clean substrate, the growth parameters can be studied. In several papers that report on nanowire position control, it has been shown that the gold dots can break up into several droplets [9,21], which resulted in several nanowires being produced from one intended gold island. Detailed studies of this phenomenon have been published for the growth of carbon nanotubes, where arrays are fabricated with the same techniques [22-23]. We have investigated this effect by systematically varying the thickness of the gold layer from 1 nm to 6 nm. Clearly for 1 nm and 2 nm (see figure 3a and 3b) the gold dots split into several droplets (probably during the anneal), catalyzing the growth of several NWs close to each other. The gold splitting can be explained by the concept of surface tension. For a small thickness, the ratio surface/volume (or diameter/thickness) ratio is large, and the islands have a pancake-like shape. Due to the surface tension the gold dots split up to form spherical particles, which will decrease the liquid/air interface. At a thickness of 4 nm (Figure 3c) still a few extra nanowires grow, however at 6 nm (Figure 3d) no splitting of the gold droplet is observed.

In order to test the quality of the nanowires grown in the arrays, we have synthesised InP/InAsP/InP heterostructured nanowires from the nanoimprinted Au islands. The optical properties of the InAsP section sensitively depend on the segment length and As/P ratio. We have applied different growth times and $AsH_3$ flows to fabricate segments with different lengths and compositions. In Figure 4a a representative tilted view SEM image is shown. The overall wire length is determined to be 5 microns. The base of the wire (up to 2 microns) is tapered, but the top part has a constant diameter. Such wire arrays are interesting for their photonic properties. We have used TEM HAADF and EDX measurements to determine the length and composition of the InAsP segments. The segment shown in Figure 4b has a length of around 22 nm and has a $In_{50}As_{37}P_{13}$ composition. The quality of these nanowire segments has been studied optically by measuring the photoluminescence at 5K. The nanowires were transferred on a $SiO_2$ substrate to measure their optical properties, as the pitch between InP nanowires is too small to optically characterize standing single nanowire. Figure 4(c) presents the power dependence of a 10 nm long $In_{50}As_{37}P_{13}$ segment in the InP nanowire. In figure 4d is shown that the exciton (1.358 eV) emission at low excitation power (linear dependence) and a biexciton line (1.354 eV) evolves at higher powers (superlinear dependence). The line width (FWHM) is 2 meV for both peaks, which is higher than previous results obtained on InP/InAsP/InP nanowires [6]. In parallel with the biexciton line, a peak appears at 1.398 eV, corresponding to the PL emission of the InP nanowire segments. We show in the inset the exciton and biexciton intensities dependence with the power, and slopes of 1 and 2 are obtained for the exciton and biexciton. These results show that these dots grown using nanoimprint technique have a good optical quality and pave the way to large scale opto-nanoelectronic applications and photonic applications.

## Conclusions

A generic process has been developed to grow arrays of (heterostructured) InP and GaP nanowires. Substrate conformal nanoimprint lithography has been successfully used to pattern gold particles on the substrate. We have shown that the preparation of the samples before growth with piranha solution in combination with a thermal anneal is a significant step to obtain perfect arrays. We are able to pattern complete 2 inch wafers with perfect nanowire arrays with excellent uniformity. This cleaning procedure is applicable to other lithography techniques as e-beam lithography, and therefore it is a generic process.

## Acknowledgements

This research was carried out under project number MC3.05243 in the framework of the strategic research program of the Materials innovation institute (M2i) (www.M2i.nl), the former Netherlands Institute of Metals Research, the FP6 NODE (015783) project, the ministry of economic affairs in the Netherlands (NanoNed). This work is part of the research program of the ''Stichting voor Fundamenteel Onderzoek der Materie (FOM)'', which is financially supported by the ''Nederlandse organisatie voor Wetenschappelijk Onderzoek (NWO)'' and is part of an industrial partnership program between Philips and FOM. The authors thank H. de Barse for SEM imaging. Correspondence and requests for materials should be addressed to E.P.A.M. Bakkers.

## References

[1]     X.F. Duan, Y. Huang, Y. Cui, J.F. Wang, C.M. Lieber, Indium phosphide nanowires as building blocks for nanoscale electronic and optoelectronic devices. *Nature*. 2001, **409**(6816), 66-69

[2]     E.D. Minot, F. Kelkensberg, M. Van Kouwen, J.A. Van Dam, L.P. Kouwenhoven, et al., Single Quantum dot nanowire LEDs. *Nano Letters*. 2007, **7**(2), 367-371

[3]     M.H.M. Van Weert, A. Helman, W. Van Den Einden, R.E. Algra, M.A. Verheijen, et al., Zinc Incorporation via the Vapor-Liquid-Solid Mechanism into InP Nanowires. *Journal of the American Chemical Society*. 2009, **131**(13), 4578-+

[4]     S.L. Diedenhofen, G. Vecchi, R. Algra, A. Lagendijk, A. Hartsuiker, et al., Broad-band and omnidirectional antireflection coating based on semiconductor nanorods. *Advanced Materials*. 2009, **21**, 973-978

[5]     R.S. Wagner, W.C. Ellis, Vapor-Liquid-Solid mechanism of single crystal growth. *Applied Physics Letters*. 1964, **4**(5), 89-90.

[6] M.H.M. van Weert, N. Akopian, U. Perinetti, M.P. van Kouwen, R.E. Algra, M.A. Verheijen, E.P.A.M. Bakkers, L.P. Kouwenhoven and V. Zwiller, Selective Excitation and Detection of Spin States in a Single Nanowire Quantum Dot *Nano Letters* 2009, **9**(5), 1989–1993.

[7] M.T. Borgström, G. Immink, B. Ketelaars, R. Algra, E.P.A.M. Bakkers, Synergetic nanowire growth. *Nature Nanotechnology*. 2007, **2**, 541-544.

[8] J. van Tilburg, R. Algra, G. Immink, M. Verheijen, E.P.A.M. Bakkers and L.P. Kouwenhoven, Surface passivated InAs/InP core/shell nanowires, *accepted for publication in Semiconductor Science and Technology*

[9] T. Mårtensson, M. Borgstrom, W. Seifert, B.J. Ohlsson, L. Samuelson, Fabrication of individually seeded nanowire arrays by vapour-liquid-solid growth. *Nanotechnology*. 2003, **14**(12), 1255-1258.

[10] L.E. Jensen, M.T. Bjork, S. Jeppesen, A.I. Persson, B.J. Ohlsson, et al., Role of Surface Diffusion in Chemical Beam Epitaxy of InAs Nanowires. *Nano Letters*. 2004, **4**(10), 1961-1964

[11] P. Mohan, J. Motohisa, T. Fukui, Controlled growth of highly uniform, axial/radial direction-defined, individually addressable InP nanowire arrays. *Nanotechnology*. 2005, **16**(12), 2903-2907.

[12] H.J. Fan, W. Lee, R. Scholz, A. Dadgar, A. Krost, et al., Arrays of vertically aligned and hexagonally arranged ZnO nanowires: a new template-directed approach. *Nanotechnology*. 2005, **16**(6), 913.

[13] H.J. Fan, B. Fuhrmann, R. Scholz, F. Syrowatka, A. Dadgar, et al., Well-ordered ZnO nanowire arrays on GaN substrate fabricated via nanosphere lithography. *Journal of Crystal Growth*. 2006, **287**(1), 34-38.

[14] T. Mårtensson, P. Carlberg, M.T. Borgström, L. Montelius, W. Seifert, et al., Nanowire arrays defined by nanoimprint lithography. *Nano Letters*. 2004, **4**(4), 699-702.

[15] Martensson T, Wagner JB, Hilner E, et al. Epitaxial growth of indium arsenide nanowires on silicon using nucleation templates formed by self-assembled organic coatings *Advanced Materials*. 2007, **19**(14), 1801-+.

[16] M.A. Verschuuren, S.F. Wuister. Imprint lithography. US patent US 2008/0011934, 17/01/08.

[17] T.W. Odom, J.C. Love, D.B. Wolfe, K.E. Paul, and G.M. Whitesides, *Langmuir* 2002, **18**, 5314-5320

[18] M. Verschuuren and H. van Sprang, 3D Photonic Structures by Sol-Gel Imprint Lithography, *Mater. Res. Soc. Symp. Proc.* Vol. 1002 1002-N03-05

[19] A.R. Clawson, Guide to references on III-V semiconductor chemical etching. *Materials Science and Engineering: R: Reports*. 2001, **31**(1-6), 1-438.

[20] G. B. Stringfellow, Organometallic Vapor-Phase Epitaxy:Theory and Practice Academic Press(New York), 1989.

[21] A.L. Roest, M.A. Verheijen, O. Wunnicke, S. Serafin, H. Wondergem, et al., Position-controlled epitaxial III-V nanowires on silicon. *Nanotechnology*. 2006, **17**(11), S271.

[22] A.V. Melechko, T.E. Mcknight, D.K. Hensley, M.A. Guillorn, A.Y. Borisevich, et al., Large-scale synthesis of arrays of high-aspect-ratio rigid vertically aligned carbon nanofibres. *Nanotechnology*. 2003, **14**(9), 1029.

[23] V.I. Merkulov, D.H. Lowndes, Y.Y. Wei, G. Eres, E. Voelkl, Patterned growth of individual and multiple vertically aligned carbon nanofibers. *Applied Physics Letters*. 2000, **76**(24), 3555-3557.

**Figures Captions:**

Table 1:

| Solution | Treatment time | InP nanowires | | |
|---|---|---|---|---|
| | | undesired | missing | misplaced |
| HNO₃ | 5 min | 10 % | <1% | <1% |
| HNO₃+HF | 5+1 min | 14 % | <1% | 7% |
| King's Water | 5 sec | 95% | 1% | 13% |
| Piranha | 30 sec | 2.5 % | <1% | <1% |

Influence of different chemical treatments on the InP nanowire growth quality. The Au layer thickness was 6 nm in all cases. The samples have been annealed at 550°C for InP just before growth. The percentage of defects (undesired, missing or misplaced nanowires) is determined from a large number (400-2000) of nanowires.

| Solution | Treatment time | GaP nanowires | | |
|---|---|---|---|---|
| | | undesired | missing | misplaced |
| 10000 Br₂/MeOH | 2 sec | 100% | 5% | 65% |
| Piranha | 5 min | 104% | <1% | 13% |
| Piranha | 3 min | 300% | <1% | <1% |
| Piranha | 1 min | 535% | <1% | <1% |
| Piranha | 1 min* | 6% | <1% | <1% |
| Piranha + HNO₃ | 1+5 min* | 9% | <1% | 6% |
| HNO₃ + Piranha | 5+1 min* | 5% | 27% | 15% |

*annealing temperature at 700°C

Table 2:

Influence of different chemical treatments on the GaP nanowire growth quality. The Au layer thickness was 6 nm in all cases. The samples have been annealed at 550°C for GaP, but samples indicated with an * have been annealed at 700°C. The percentage of defects (undesired, missing or misplaced nanowires) is determined from a large number (400-2000) of nanowires.

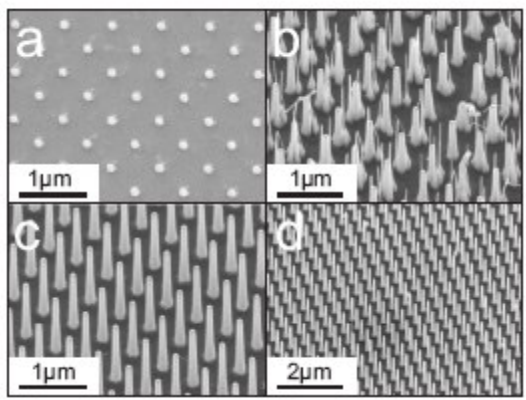

Figure 1:

InP nanowire arrays by soft nano imprint lithography. (**a**) Pattern of Au dots after lift-off process, prior to the growth of nanowires. (**b**) Nanowire array without suitable cleaning step prior to growth resulting in extra "grass-like" wire growth between the patterned array. (**c**) and (**d**) Nanowire array with a piranha cleaning step and 550˚C anneal before growth .

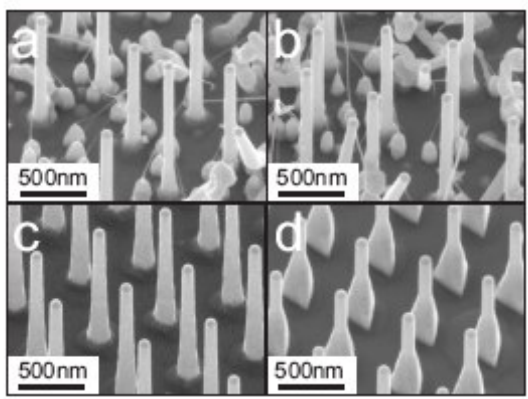

Figure 2:

InP nanowire arrays by surface conformal imprint lithography. Influence of annealing steps prior to growth. (**a**) No annealing (**b**) 420°C anneal step (**c**) 550°C anneal step (**d**) 700°C anneal step. All samples are pre-treated with a piranha etch.

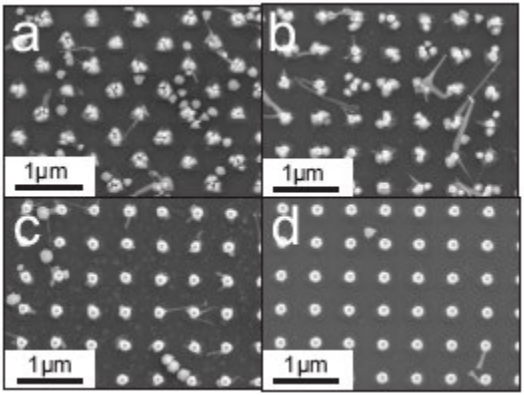

Figure 3:

Influence of Au thickness deposited on the imprinted arrays on the quality of nanowire growth. (**a**) 1nm Au (**b**) 2nm Au (**c**) 4nm Au (**d**) 6nm Au. All samples are pre-treated with an anneal step at 550 °C and a piranha etch.

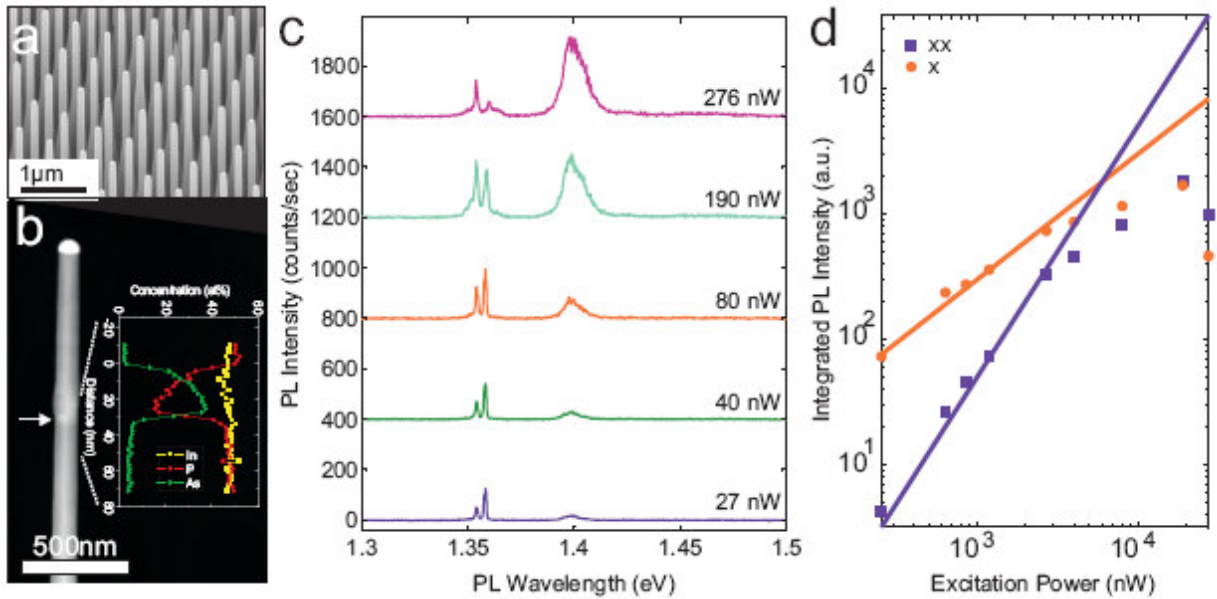

Figure 4:

Heterostructured InP-InAsP-InP nanowires. (a) SEM image viewing the array of heterostructured nanowires (b) HAADF TEM image showing the InAsP segment (bright) indicated by an arrow in the InP nanowires (dark). The inset shows the As concentration along the nanowire axis. The increased As concentration clearly shows the InAsP segment. Different As concentrations are observed before and after the InAsP segment. This is due to an As shell around the first InP part segment. (c) Photoluminescence data showing the quality of the dot emission grown in a patterned nanowire array.

Supplementary Figure captions:

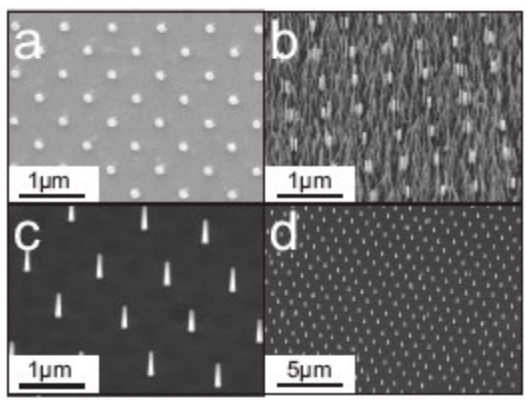

Supp. Figure 1:

GaP nanowire array by soft nano imprint lithography. (**a**) Pattern of Au dots after lift-off process, prior to the growth of nanowires. (**b**) Nanowire array without suitable cleaning step prior to growth resulting in extra "grass-like" wire growth between the patterned array. These extra undesired nanowires occur by using King's Water, $HNO_3$, HF, or $Br_2$/MeOH. (**c**) and (**d**) Nanowire array with a piranha cleaning step and 700°C anneal before growth

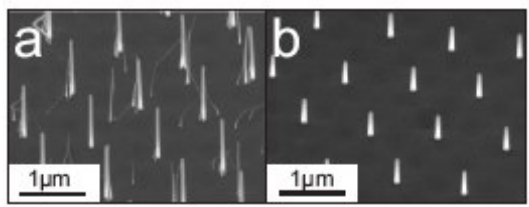

Supp. Figure 2:

Influence of annealing steps prior to growth. (**a**) 550°C anneal step (**b**) 700°C anneal step. All samples are pre-treated with a piranha etch.